\documentclass[fleqn,10pt]{wlscirep}
\usepackage[utf8]{inputenc}
\usepackage[T1]{fontenc}
\usepackage{lineno}
\usepackage{xcolor}
\usepackage{subcaption}
\usepackage{multirow}

\title{ORBITAAL: A Temporal Graph Dataset of Bitcoin Entity-Entity Transactions}

\author[1,*,$\dag$]{Célestin Coquidé}
\author[1,$\dag$]{Rémy Cazabet}
\affil[1]{LIRIS UMR 5205 CNRS / Universite Claude Bernard Lyon 1 / INSA Lyon / Université Lumière Lyon 2 / École Centrale de Lyon, 69100 Villeurbane, France.}
\affil[*]{corresponding author(s): Célestin Coquidé (celestin.coquide@liris.cnrs.fr}

\affil[$\dag$]{these authors contributed equally to this work}

\begin{abstract}
Research on Bitcoin (BTC) transactions is a matter of interest for both economic and network science fields. Although this cryptocurrency is based on a decentralized system, making transaction details freely accessible, making raw blockchain data analyzable is not straightforward due to the Bitcoin protocol specificity and data richness. To address the need for an accessible dataset, we present \textit{ORBITAAL}, the first comprehensive dataset based on temporal graph formalism. The dataset covers all Bitcoin transactions from January 2009 to January 2021. \textit{ORBITAAL} provides temporal graph representations of entity-entity transaction networks, snapshots and stream graph. Each transaction value is given in Bitcoin and US dollar regarding daily-based conversion rate. This dataset also provides details on entities such as their global BTC balance and associated public addresses.

\end{abstract}
\begin{document}

\flushbottom
\maketitle

\thispagestyle{empty}

\section*{Background \& Summary}
Bitcoin (BTC) is the first modern cryptocurrency. It was proposed in 2008 by a person or a group of person known as Satoshi Nakamoto, implementing a protocol described in a white paper \cite{nakamoto_bitcoin_2008}. Unlike economic activities issued by banks and other financial institutions, all details about past Bitcoin transactions are freely accessible to anyone with the skill to access it, since all the transactions are recorded in the so-called blockchain. Since the first BTC transaction happened in January 2009, the Bitcoin ecosystem has brought interest as a source of investment, speculation, and is also used to buy and sell goods and services such as dark market and gambling platforms. In order to perform BTC transactions, users use pairs of public and private keys instead of bank accounts. This pseudonymity motivates cybercriminals to use such a system.

While using networks to represent Bitcoin transactions seems natural, the raw data stored in the blockchain is not adapted for an efficient representation as such. Indeed, due to the design of the Bitcoin protocol, each transaction can involve multiple inputs and multiple outputs, thus requiring a representation as a directed and weighted hypergraph, that is complex and inconvenient. Even worse, the nodes of this network would be Bitcoin addresses, which are significantly different from Bitcoin users. The usual approach in the literature thus consists of using heuristics and other strategies to identify the multiple addresses of a single user, allowing to move from an address-to-address network to a user-to-user one.

To the best of our knowledge, no complete Bitcoin dataset designed for graph analysis has been published in the scientific literature. Among existing datasets, some only contain samples of data, while others (e.g. \cite{emery2021full}) only provide raw data, requiring researchers to perform multiple pre-processing steps, complex due to the massive scale of the data. 
For instance, the dataset by Kondor et al.\cite{kondor_bitcoin_2020} provides rich information on the blockchain and transaction level, but requires a technical transformation of the data to obtain exploitable user-graph data. Elliptic dataset\cite{elmougy_demystifying_2023, ellipticsolution} is the most popular network-based dataset, used in machine learning studies, but only considers address-to-address transactions. Elliptic also includes only 200k transactions representing a very small fraction of Bitcoin transactions. One of the most common way researchers use to access complete Bitcoin Data is through specialized open source data extraction software\cite{kalodner_blocksci_2020, liu_deciphering_2022}. However, the extracted data still requires processing to be transformed into analyzable networks. Furthermore, the most used of them, BlockSci\cite{kalodner_blocksci_2020} is no longer supported since 2020. There are also online cryptocurrency explorer and analytics services that don't provide any dataset permitting the construction of transactions network\cite{blockchain_com, coinmetricservice, ellipticsolution}.

The need for a complete and comprehensive Bitcoin dataset for graph analysis goes beyond the research on cryptocurrency and the economic relations between users. Indeed, the temporal aspect of Bitcoin transactions and the large number of users and transactions make it a perfect candidate for temporal graph analysis, especially for the development of large network analysis tools and adapted algorithms for big data analysis.

Aware of the importance of the economic network of Bitcoin and the necessity for a comprehensive dataset, we propose \textit{ORBITAAL}\cite{orbitaal} (cOmpRehensive BItcoin daTaset for temorAl grAph anaLysis), a readily analyzable dataset, structured in a standard formalism, i.e., a temporal graph. \textit{ORBITAAL} requires no specific Bitcoin-related knowledge to be used, and covers the period 2009 to 2021, representing 13 years of Bitcoin transactions. It is also the first dataset proposing two temporal graph representations: the stream graph, as described in \cite{latapy_stream_2018}, and snapshots using different timescales, namely year, month, day and hour. All transactions present in the dataset are expressed in both BTC and daily-based US dollar (USD) converted values. In addition to the snapshots and stream graph, we also provide the list of all Bitcoin users with their name, first and last period of activities, and global BTC balance. Since these users have been inferred from aggregation techniques and set of addresses found in {\url{walletexplorer.com}}, we also give the list of all users' pool of addresses.

\section*{Methods}
Due to the decentralized nature of cryptocurrencies such as Bitcoin, the data stored in their blockchain is accessible by anyone. Therefore, researchers can access it without any cost and limitations. The data stored in the Bitcoin blockchain is rich,  containing information such as public addresses, transaction hash, amount of BTC exchanged, timestamp of the transactions, fees paid by senders, etc.  We describe in this section the methods used to build \textit{ORBITAAL} from the extraction of the blockchain to the construction of the temporal graphs.

\subsection*{From binary data to human readable blockchain contents}
The first step of the data collection was to download the full content of the Blockchain. To do so, one has to install the Bitcoin Core software, and to launch the synchronisation of the full Blockchain. At the time of download, the full weight of the blockchain was about 207 GB.

The second step consists in transforming the raw data, provided in an efficient Binary format, into an easily parsable and interpretable format. 
We used Bitcoin-etl python library\footnote{\url{https://github.com/blockchain-etl/bitcoin-etl}} to transform this raw data into json format. After transformation, the full data weighs 1.6 TB.

The next step consists in enriching the data with input Bitcoin addresses. Indeed, in Bitcoin, each input is an output from a previous transaction. In the original format, inputs are thus identified by a pair $(TxID, N)$, with $TxID$ the unique hash of the input transaction, and $N$ an integer identifying which output of $TxID$ is used as input. Since we need to know the identity of the sender, we retrieve the Bitcoin address information for each input of each transaction.

\subsection*{Address clustering}
In the \textit{ORBITAAL} dataset, Bitcoin users are first inferred using the common-input heuristic \cite{harrigan_unreasonable_2016}. This task has been performed efficiently by using a graph approach: 1)We first create a graph in which each node is an address, 2)For each transaction, we add an edge between the input address in position $x$ and the one in position $x+1$, 3)finally, we compute the connected components of this graph using an efficient graph framework, namely NetworKit\cite{staudt2016networkit}. Each connected component constitutes a group of addresses considered as belonging to the same user. This heuristic has been repeatedly considered efficient in the literature\cite{harrigan_unreasonable_2016}, despite its known limits\cite{moser2017price}. The results of user identification should therefore always be interpreted with caution, although in phase with the state of the art.
Additionally to this heuristic, we also use a well-known website, WalletExplorer\cite{walletex}, that provides a database linking Bitcoin addresses to the real identity of their owner. We use it for two tasks: 1)Enrich the data with 360 user identities, including Exchange platforms, mining pools, gambling services, etc. and 2) Further merge clusters of addresses that are known to belong to the same user based on this external knowledge.

\subsection*{User to user transactions}
After performing address clustering, each Bitcoin transaction becomes a one-to-one or one-to-many transaction: all addresses in input by definition belong to the same user, but addresses in output might belong to 1)the same user (self-spent), 2)one different user, 3)multiple different users, including or not the sender. We convert these Bitcoin transactions into simpler, one-to-one payments: for a Bitcoin transaction from user $u_1$ having $p$ different outputs to users $(u_2, u_3, \dots,u_{1+p})$., we create a corresponding number of payments $(u_1,u_2), (u_1,u_3),\dots,(u_{1},u_{p+1})$. The amount associated with this payment is the sum of all payments from the source to the destination at that particular timestep. 

\begin{figure}[htbp]
    \begin{center}
        \begin{subfigure}[b]{0.4\textwidth}
        \centering
        \raisebox{5pt}{\includegraphics[width=\textwidth]{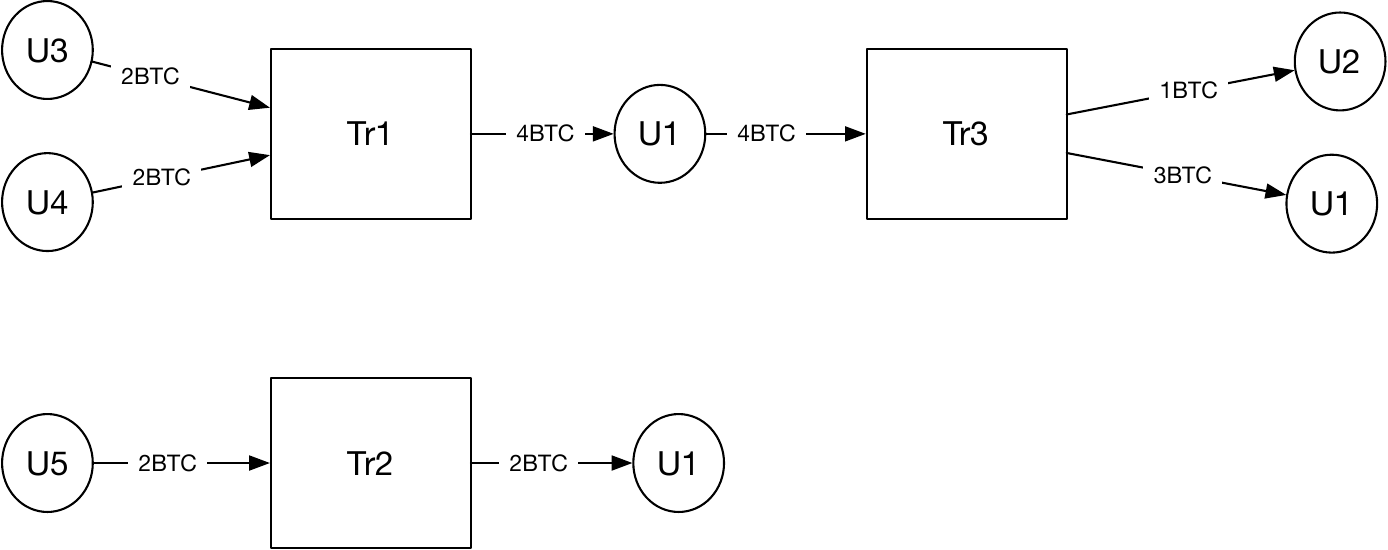}}
        \caption{Blockchain representation. Transactions occur in this order: Tr1, Tr2, Tr3.}
        \label{fig:figa}
    \end{subfigure}
    \begin{subfigure}[b]{0.4\textwidth}
        \centering
        \raisebox{30pt}{\includegraphics[width=\textwidth]{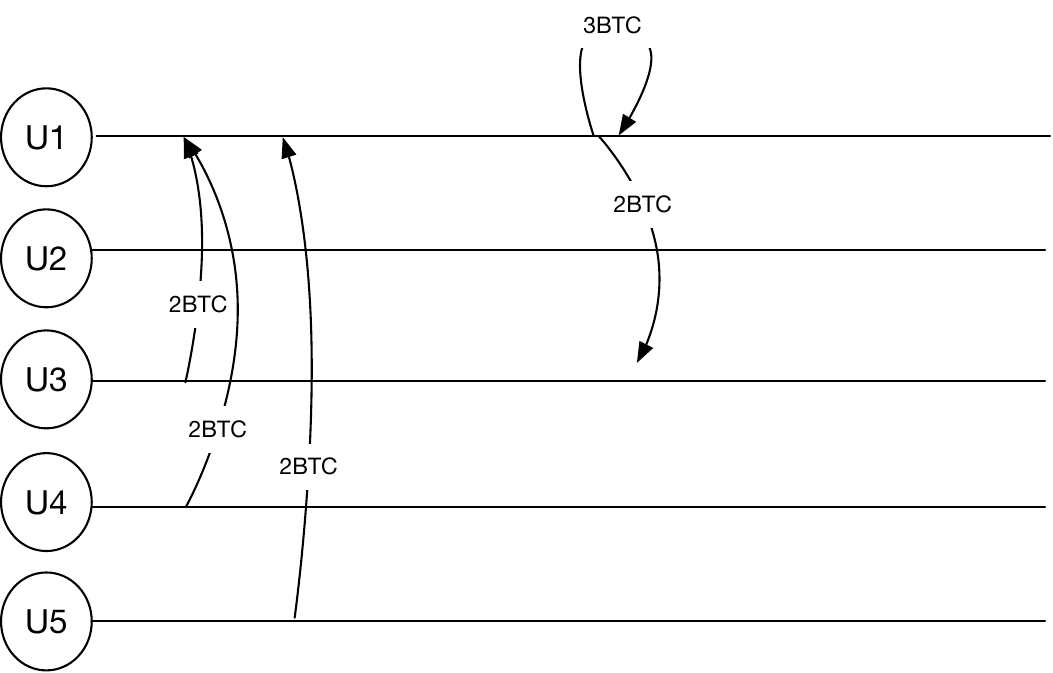}}
        \caption{Stream graph representation}
        \label{fig:figb}
    \end{subfigure}
    \end{center}
    \caption{Illustrations of the change of representation between the original Blockchain Data and our Stream Graph representation. Note the loss of information: the original data (left panel) keeps track of the exact flow, e.g., that coins sent by U1 to U2 come from U3 and U4, not U5. On the contrary, in the user-graph representation (right panel), this information is lost. However, in most cases, this information can be considered as noise, and the user-network representation is more faithful to how users behave in the network.   }
    \label{fig:informationLoss}
\end{figure}

Note that the transformation from an address network to a user network, although simplifying greatly the data, comes at the price of some information loss compared with the raw data. Namely, there are two sources of information loss: 1)We lose the details of the addresses used, which might contain useful details such as the fragmentation of coins in multiple accounts, etc. 2) We lose the information of multiple payments between the same actors occurring at the same time. However, we think that most of this information constitutes noise, which is not relevant to the analysis. To take a concrete example (see Fig. \ref{fig:informationLoss}), the raw data contains all details about sources and destinations, e.g., if a user $u_1$ makes a transaction to pay user $u_2$, one could observe that the transaction has 2 inputs, one coming from user $u_3$ and another from user $u_4$, although user $u_1$ also received coins from $u_5$. One could therefore trace coins from $u_3$/$u_4$ to $u_2$ through $u_1$, and not from $u_5$ to $u_2$ through $u_1$. When performing the user-aggregation, this information is lost, and we only see that $u_1$ received coins from $u_3$,$u_4$ and $u_5$ before making the payments to $u_2$. However, it is very likely that those details were in fact misleading: if for instance $u_1$ is using a wallet software to manage its activities, they are not aware of the details of those transactions as described above, the wallet automatically chooses the right sources to make its payment while minimizing the fees to pay. In most cases, we can therefore assume that this information loss is acceptable, or even beneficial to the interpretation of user activity. 

\subsection*{Mining node}
In Bitcoin and many other cyptocurrencies, the validation of transactions and their inclusion in the blockchain is performed by specific users, known as miners. When a set of transactions (bloc) is mined and added to the blockchain, the actor who solved the required cryptographic puzzle earns rewards in the form of newly minted Bitcoins. Additionally, the miner also receive payments from senders in the form of transaction fees. This is because the Bitcoin protocol has a fixed, limited bandwidth, requiring users to compete for their transactions to be included. Users spending Bitcoin have thus to pay fees, acting as incentives to miners for including these transactions in the blockchain, as Miners can choose which transaction they want to include in the next block. To represent these fee transactions while keeping a graph-only representation, we introduce a formalism using a dedicated node, called Miner-Node. Its outgoing transactions encode miners' rewards, and its incoming transactions for paid fees.

At this stage, we have a complete dataset of user-user transactions. To make it easily reusable, we convert it into two standard temporal graph formats, stream graph and snapshot sequences.

\subsection*{Stream graph construction}
Stream graph formalism was proposed by Latapy et al. in 2018\cite{latapy_stream_2018}. This temporal graph representation is the most natural for data composed of interactions, such as payments between users. We note $G=(T,V,W,E)$ the stream graph consisting in a set of times $T$ (every instant at which at least a transaction has been added to the blockchain), a set of vertices $V$ corresponding to users that have been involved in at least one transaction, the set of temporal nodes $W$, pairs $(u,t)$, $u\in V$,$t\in T$, representing the existence of node $u$ at time $t$, and the set of temporal links $E$, 3-tuples $(u,v,t)$ corresponding respectively to the spender, receiver and time of the payment. The mining dedicated node is also present in the stream graph and handles both the paid fees and mining rewards. 

In the stream graph, edges are directed, and information is associated with nodes (name, list of addresses) and edges (BTC amount, US Dollar value).

\begin{figure}[ht]
    \begin{center}
        \begin{subfigure}[b]{0.4\textwidth}
        \centering
        \raisebox{5pt}{\includegraphics[width=\textwidth]{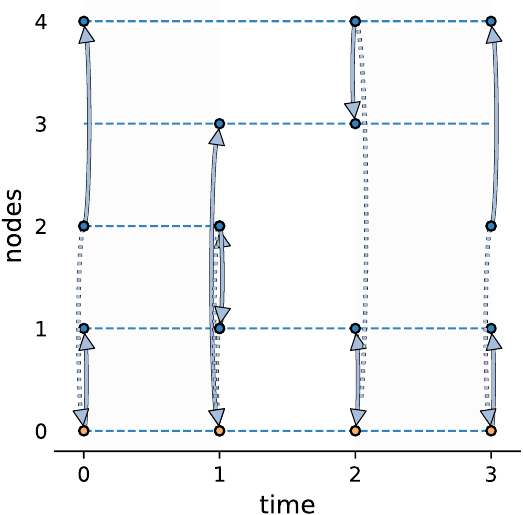}}
        \caption{Stream graph}
        \label{fig:sg}
    \end{subfigure}
    \begin{subfigure}[b]{0.4\textwidth}
        \centering
        \raisebox{30pt}{\includegraphics[width=\textwidth]{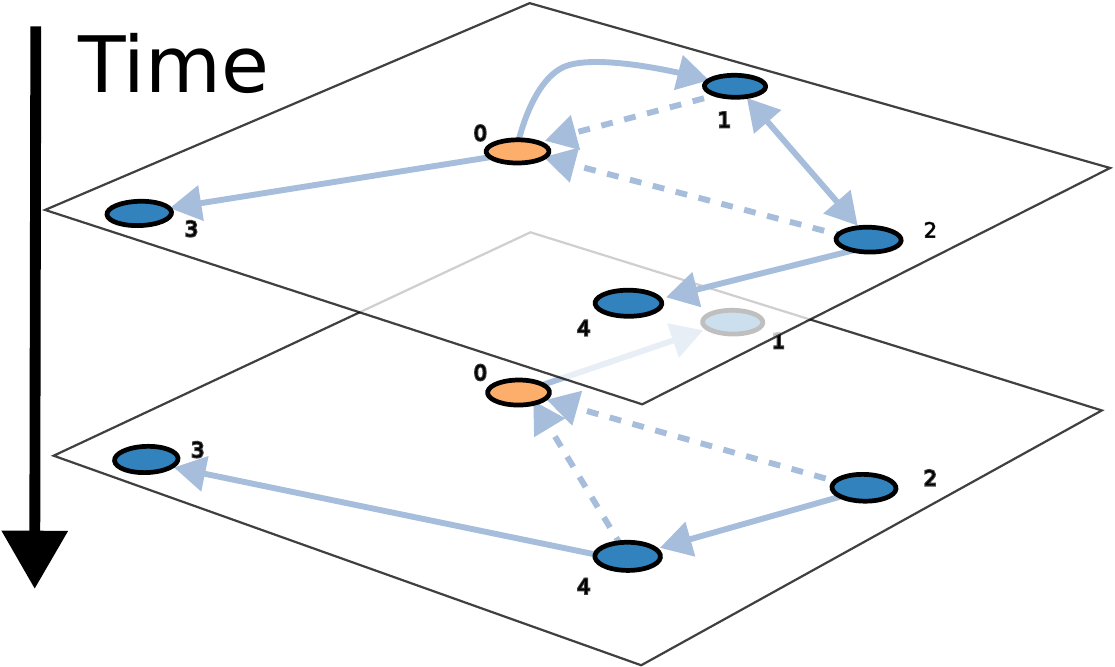}}
        \caption{Snapshots}
        \label{fig:sn}
    \end{subfigure}
    \end{center}
    \caption{Illustrations of snapshot (left panel) and stream graph (right panel) temporal network representation of Bitcoin transfers between users. The orange-colored node represents the mining node and dotted links are associated with fee transactions.}
    \label{fig:toynet}
\end{figure}

This Bitcoin user transaction network represents all BTC transactions at the finest timescale ---being $\approx 10$ minutes, the average time between two validated blocs. An illustration of a stream graph representing Bitcoin transactions is given in Fig.~\ref{fig:toynet} panel (a).

\subsection*{Snapshot construction}
Although the stream graph representation is the most natural for this data, it is not the most common formalism to work with temporal graphs. Indeed, most existing algorithms and tools rather expect a representation as sequences of static graphs, known as snapshots. To facilitate reusability, we also provide the data as snapshot representations, aggregated at typical timescales.
Let us note $\tau$ a set of discrete times or continuous period of times, we have $G_{\tau}(E,V)$ the snapshot covering period $\tau$ with $V$ the set of Bitcoin users performing at least one transaction in $\tau$ and $E$ the set of directed links representing the direction of Bitcoin from spenders to receivers. The weight of each link is the sum of Bitcoins exchanged over $\tau$. As presented before, a dedicated mining node is added to these networks with specific ID $0$ such that the link $0 \rightarrow m $ and $s \rightarrow 0 $ represent respectively Bitcoin earned by miners $m$ and fees paid by spender $s$. A toy network is presented in Fig.~\ref{fig:toynet} panel (b).

We provide different temporal scales for Bitcoin user transaction snapshots. The largest encodes the snapshot covering the whole period of data (13 years), other resolutions are associated with year, month, day, and hour scales.

\subsection*{Additional information related to users and addresses}
Additionally to the network representation, we also provide important data about users' life cycles. In the Bitcoin protocol, there is no user creation or removal from the system, one only observes them when they are involved in a transaction. For each user, we provide a period of existence corresponding to the period between its first apparition, and the latest time at which its balance was positive ---or the end of the period covered by the dataset if it is still positive at that time. We also provide each user's list of associated public addresses (list of address aggregates), their identity as known from WalletExplorer if available, and their final BTC balance. The BTC balance is measured as the difference between the total quantity of BTC spent and received over the whole period.

\section*{Data Records}
\textit{ORBITAAL} dataset\cite{orbitaal} is available on the Zenodo open repository at {\url{https://doi.org/10.5281/zenodo.12581515}}. 
\subsection*{Dataset contents}
The files containing the main ORBITAAL dataset are the following:
\begin{itemize}
    \item \textbf{orbitaal-stream\_graph.tar.gz} contains the stream graph representation, split in one file per year.
    \item \textbf{orbitaal-snapshot-all.tar.gz }contains a single static graph, corresponding to the total cumulated graph.
    \item \textbf{orbitaal-snapshot-year.tar.gz, orbitaal-snapshot-month.tar.gz, orbitaal-snapshot-day.tar.gz} and  \textbf{orbitaal-snapshot-hour.tar.gz }contain respectively snapshots aggregations every year, month, day and hour. Each file is a compressed archive, containing one graph file per aggregation window.
    \item \textbf{orbitaal-nodetable.tar.gz} contains the node details, i.e., for each user ID, its estimated lifetime period, final balance, and list of Bitcoin addresses. The IDs used are coherent over each network representation.

\end{itemize}

Additionally to the data files, the repository includes a Readme file in MarkDown and 4 short CSV files representing transactions at the day resolution before and after the Bitcoin halving that occurred in July 2016. These CSV files can serve as quick-starting examples to get used to the dataset, without the difficulty of handling large data. Table~\ref{tab:record1} gives details on the 12 archives present in the dataset. We can note that the weights of our graph representations are much smaller than the original files extracted from the blockchain. This is because we removed a lot of information not directly related to transactions (e.g., transaction hashs), concentrating on the essential transactional aspect of the Bitcoin data.

\subsection*{Format}
Each graph data is provided in \textit{.parquet} format, which is an efficient tabular data storage, similar to .\textit{csv} but offering better performance: 1)it uses less space on disk, 2)retains original data types, and 3)allows loading only some of the columns to save memory space. .\textit{parquet} is a standard format that can be directly loaded in libraries such as Pandas\cite{pandas} or Pyspark\cite{pyspark}.

Each network is provided in an enriched edge list format, i.e., each row contains an edge corresponding to a payment (or a summary of payments over a period), described by source and destination nodes ID, a timestamp in the case of stream graphs, and values in BTC and US Dollar.

\section*{Technical Validation}
This section presents experiments demonstrating that the \textit{ORBITAAL} dataset\cite{orbitaal} captures accurately Bitcoin transaction data, and can be analyzed in a temporal graph formalism. 
\subsection*{Sanity check}

In order to assert the validity of our data processing steps, we compare some key data descriptors with an external ground truth. We use as reference \textit{blockchain.com}\cite{blockchain_com}, a popular cryptocurrency explorer, proving various statistics about the Bitcoin blockchain. Global measures are based on the stream graph data, the other ones being constructed from it. To compare the results with the ground truth, we use the relative error $\delta_{r}$, with $\delta_{r} \in [-1,\infty]$, defined as 
\begin{equation}
    \delta_{r}(v_{1},v_{2}) = \frac{v_{1}-v_{2}}{v_{2}}    
\end{equation}
where $v_{1}$ (resp. $v_{2}$) is the measure obtained from \textit{ORBITAAL} (dataset of reference). 
A value close to 0 means an exact similarity while a positive (resp. negative) value of $\delta_{r}$ indicates higher accuracy for \textit{ORBITAAL} (dataset of reference).
\newline

\begin{figure}[h]
    \centering
    \includegraphics[width=\textwidth]{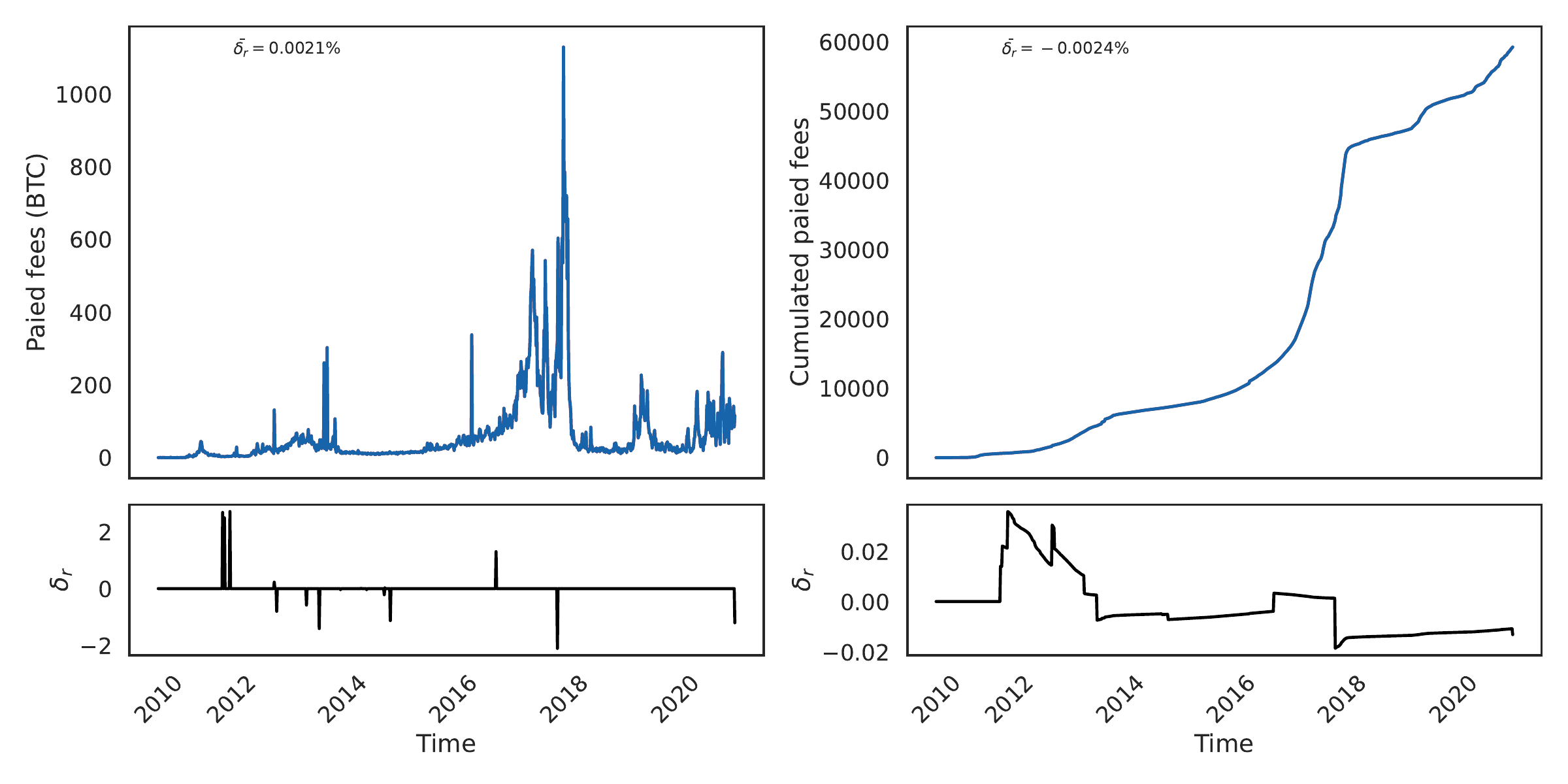}
    \caption{Comparing the daily paid fees (top left panel) and its cumulative (top right panel) obtained from \textit{ORBITAAL} dataset (blue) with data from \textit{blockchain.com} (red). The bottom panels represent the relative error between both datasets such that positive (resp. negative) $\delta_r$ value indicates a higher accuracy for \textit{ORBITAAL} (the dataset of reference.}
    \label{fig:sanityfees}
\end{figure}

The first Bitcoin feature we consider is the overall amount of transaction fees paid by all users daily. 
In our dataset, this information is captured by transactions to and from the dedicated mining node. Fig.~\ref{fig:sanityfees} shows strong similarities between datasets with an average relative error $\bar{\delta_{r}} = 0.0021\%$. This means that our representation is faithful to the original data.

\begin{figure}[h]
    \centering
    \includegraphics[width=\textwidth]{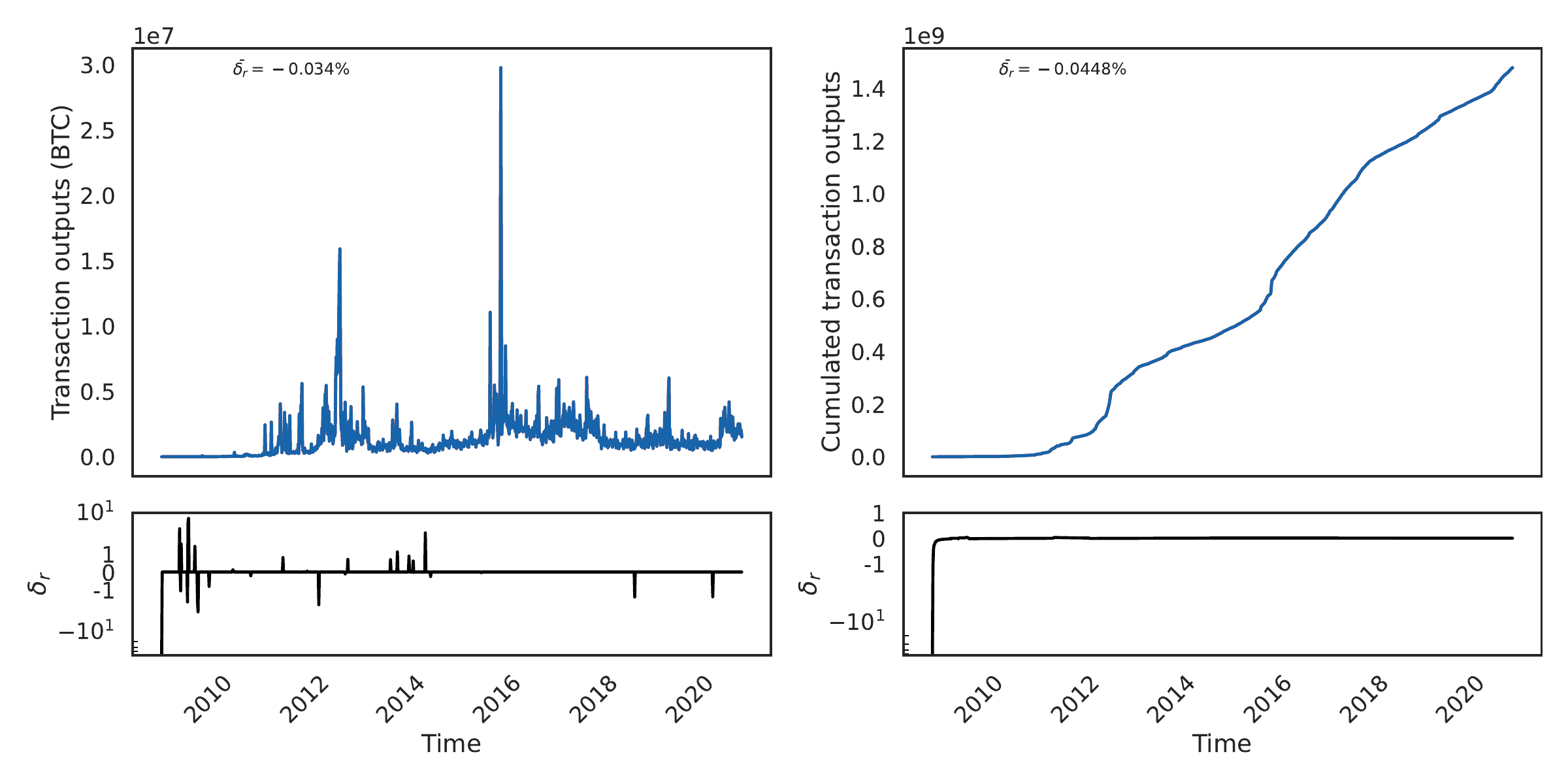}    
    \caption{Comparing the daily total transaction outputs in BTC (top left panel) and its cumulative (top right panel) obtained from \textit{ORBITAAL} dataset (blue) with data from \textit{blockchain.com} (red). The bottom panels represent the relative error between both datasets such that positive (resp. negative) $\delta_r$ value indicates a higher accuracy for \textit{ORBITAAL} (the dataset of reference.}
    \label{fig:sanitytroutputs}
\end{figure}

The second global measure compared in this experiment is the total transactions' outputs in BTC representing all exchanged Bitcoin ---not including transaction fees. Its comparative analysis with the dataset of reference is presented in Fig.~\ref{fig:sanitytroutputs}. We observe that our dataset is faithful to the reference, with $\bar{\delta_{r}} = -0.034\%$. Results related to the 2009-2010 period indicate some differences, due to non-standard transactions that were common in Bitcoin's early period. 

\begin{figure}[h]
    \centering
    \includegraphics[width=\textwidth]{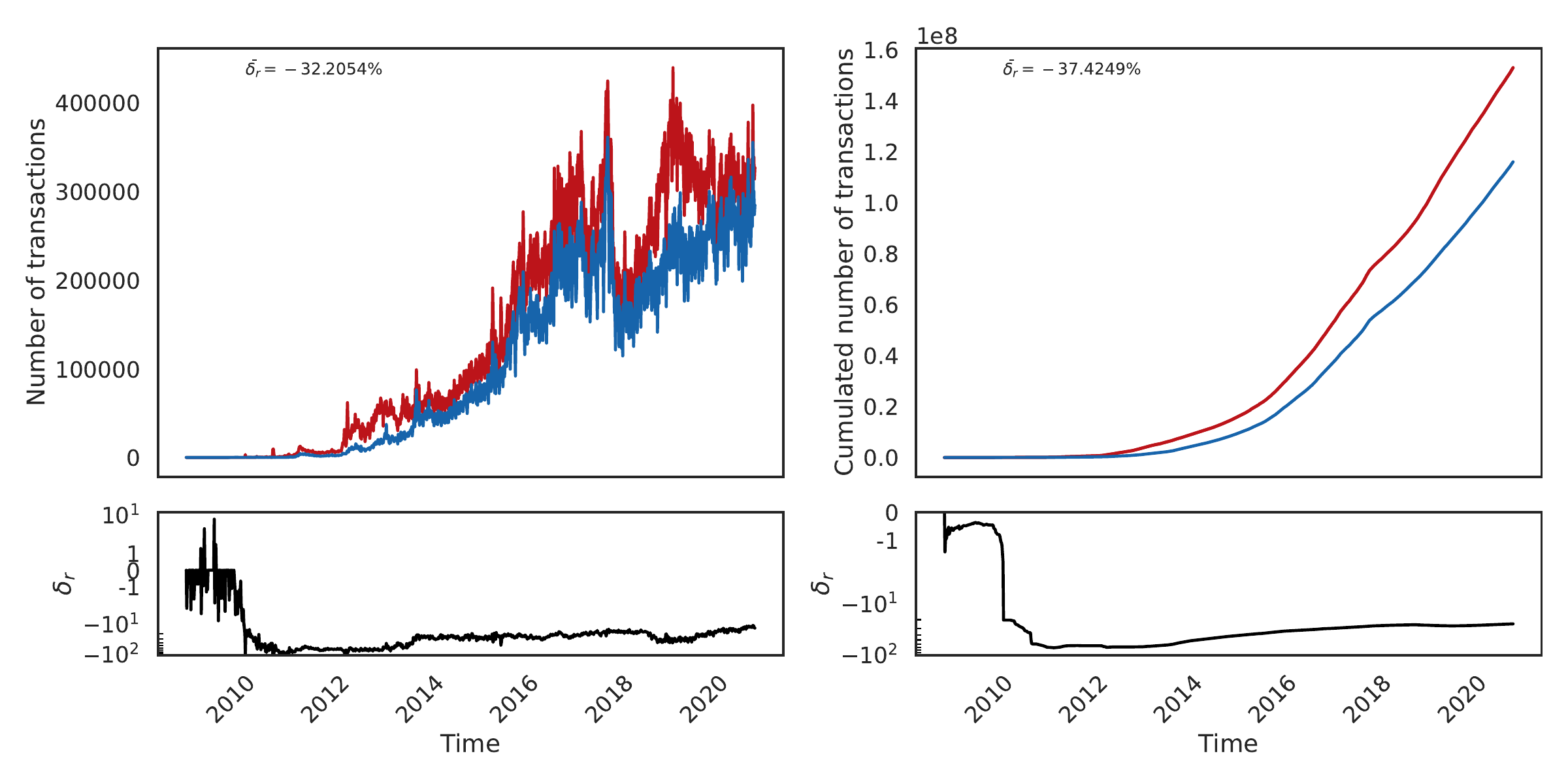}    
    \caption{Comparing the daily number of distinct transactions (top left panel) and its cumulative (top right panel) obtained from \textit{ORBITAAL} dataset (blue) with data from \textit{blockchain.com} (red). The bottom panels represent the relative error between both datasets such that positive (resp. negative) $\delta_r$ value indicates a higher accuracy for \textit{ORBITAAL} (the dataset of reference.}
    \label{fig:sanitytrcounts}
\end{figure}

The last sanity check experiment consists in comparing the number of distinct transaction outputs. This measure is directly related to the number of links in temporal graphs provided by \textit{ORBITAAL}. Since our dataset provides user-to-user transaction networks, we expect differences with the dataset of reference giving details at the  level of address-to-address. Results show that our dataset seems accurate since the relative error is stable ($\bar{\delta_{r}}=-32\%$) indicating that differences come mostly from address clustering (see Fig.~\ref{fig:sanitytrcounts}).

Based on the sanity check experiments performed on important Bitcoin features, we conclude that \textit{ORBITAAL} dataset provides accurate Bitcoin transaction networks with no significant lack of information compared with data that one can find in reference commercial services.

\subsection*{Structural analyses of Bitcoin transaction temporal graphs}
This experiment aims to show that \textit{ORBITAAL} dataset can be used for temporal graph analysis and can be used to infer changes in Bitcoin activities through network structural changes. We analyze changes in the transaction graph structure, after removing transactions representing fees.  
Snapshots are static networks, their structure can thus be investigated with standard network metrics and analysis tools. Analyzing stream graphs is less straightforward, as stream graph formalism implies new definitions and metrics. We briefly introduce the most common ones, below, following \cite{latapy_stream_2018}. 

\subsubsection*{Node contribution}
In a stream graph, a living period $\tau_{i}$ is associated with a node $i$ such that for every time instants $t \in \tau_{i}$, the doublet $(i,t)$ exists in the set of temporal nodes $W$. We define contribution of the node $i$, noted $n_{i}$, as the living period of $i$ expressed as a fraction of all time instants ($\tau$), we have
\begin{equation}
n_{i} = \frac{|\tau_{i}|}{|\tau|}
\end{equation}
with $|x|$ the size of the set $x$. We also define $n=\sum_{i\in V}n_{i}$, the total node contribution, with $V$ being the set of nodes (all users involved in at least one transaction).
\subsubsection*{Node degree}
The node degree informs on nodes' local neighborhood, which, in the context of Bitcoin transaction networks, is a measure of the transaction network associated with any user and its economic partners. In the case of directed static networks, such as snapshots, we have two measures of degree, the node's incoming degree and outgoing degree. While the first is defined as the number of incoming edges, the latter is associated with the number of outgoing edges. In the context of stream graphs, there are multiple definitions for node degree, each of them leading to different interpretations. We consider an intuitive adaptation of node degree for stream graph such that the incoming degree $k_{in}$, and outgoing degree $k_{out}$, of a node is the number of incoming, respectively outgoing, transactions per unit of time during this node living period. We have
\begin{equation}
    k_{in}(j)=\frac{Ntr_{in}(j)}{|\tau_{j}|}
\end{equation}
with $Ntr_{in}(j)$ the number of transactions $j$ is an output of.
\begin{equation}
    k_{out}(j)=\frac{Ntr_{out}(j)}{|\tau_{j}|}
\end{equation}
with $Ntr_{out}(j)$ the number of transactions $j$ is an input of.
We note $\bar{k_{in}}$ and $\bar{k_{out}}$ the incoming and outgoing degree averaged over all nodes. While these two measures are identical in the case of snapshots, they are not with stream graphs.

\subsubsection*{Strongly connected component}
A strongly connected component (SCC) of a directed network is defined as a partition of the network (sub-network) such that any node composing it can be reached from each other node that is part of it. In this experiment, we consider the largest SCC which is the SCC composed by the highest number of nodes. Since the largest SCC captures the large connected part of the network, analyzing its evolution is a relevant indicator for structural changes. Although there is a stream graph equivalent of SCC, the large size of Bitcoin transaction graphs makes its extraction difficult. Therefore we believe that \textit{ORBITAAL} is a perfect candidate for large network analysis tools development and considering stream graph equivalent for SCC is part of further researches.

\subsubsection*{Diameter and average shortest path}
Let $i$ and $j$ be two nodes from a directed network, the path from $i$ to $j$ is the sequence of links used to reach $j$ from node $i$. The shortest path between these two nodes will be the path presenting the lowest length (number of edges composing the path sequence). We can define two global distance metrics from it, the diameter which is the highest shortest path between any pair of nodes, and the average shortest path. In the context of Bitcoin transaction network, these two metrics give access to typical distances between any users, in terms of the number of transactions. The equivalent for stream graphs are temporal paths and distances, but they were not considered in this article due to the lack of scalable implementations. 

\subsubsection*{Results}
\begin{figure}[h]
    \centering
    \includegraphics[width=\textwidth]{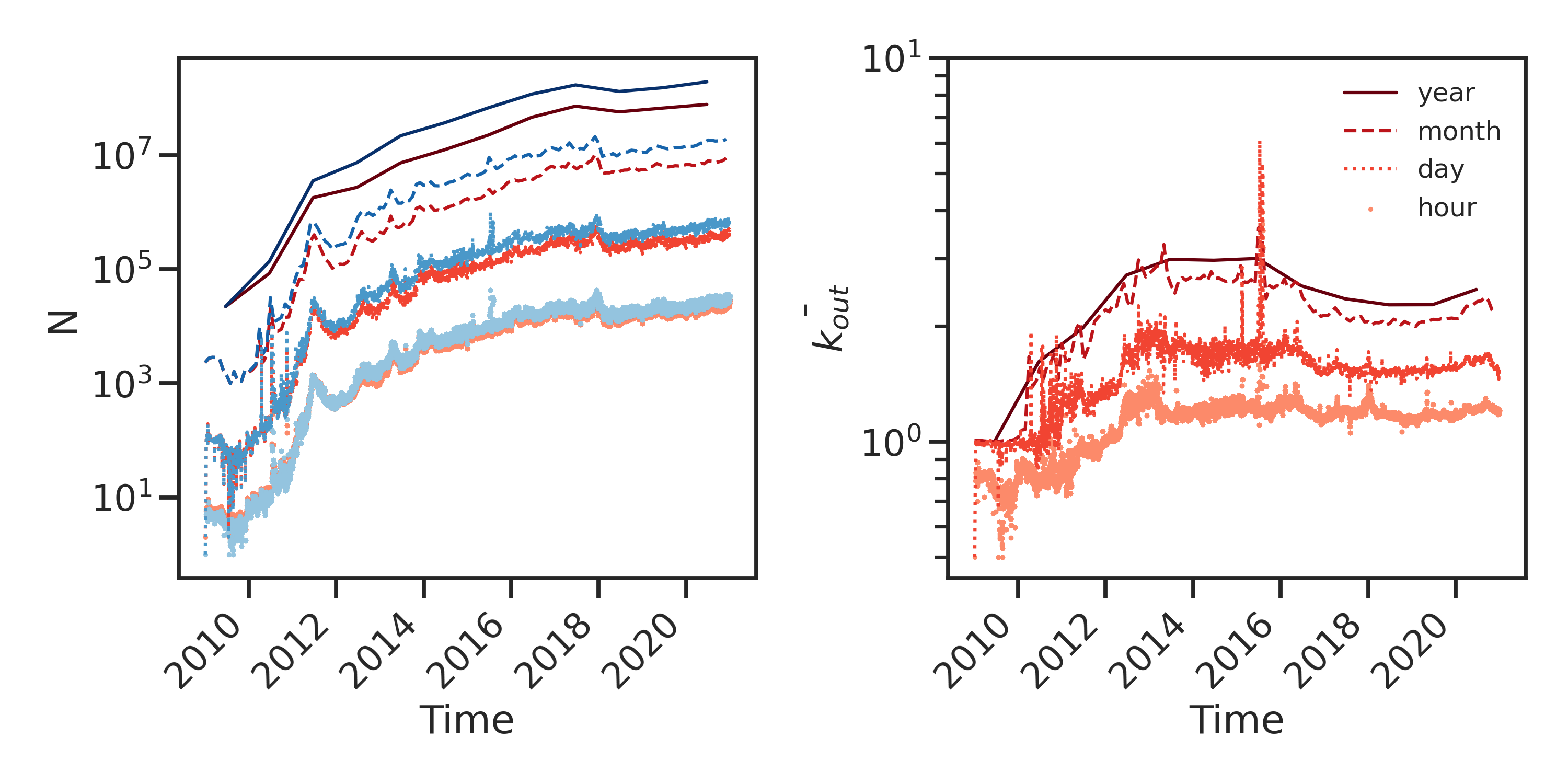}
    \includegraphics[width=\textwidth]{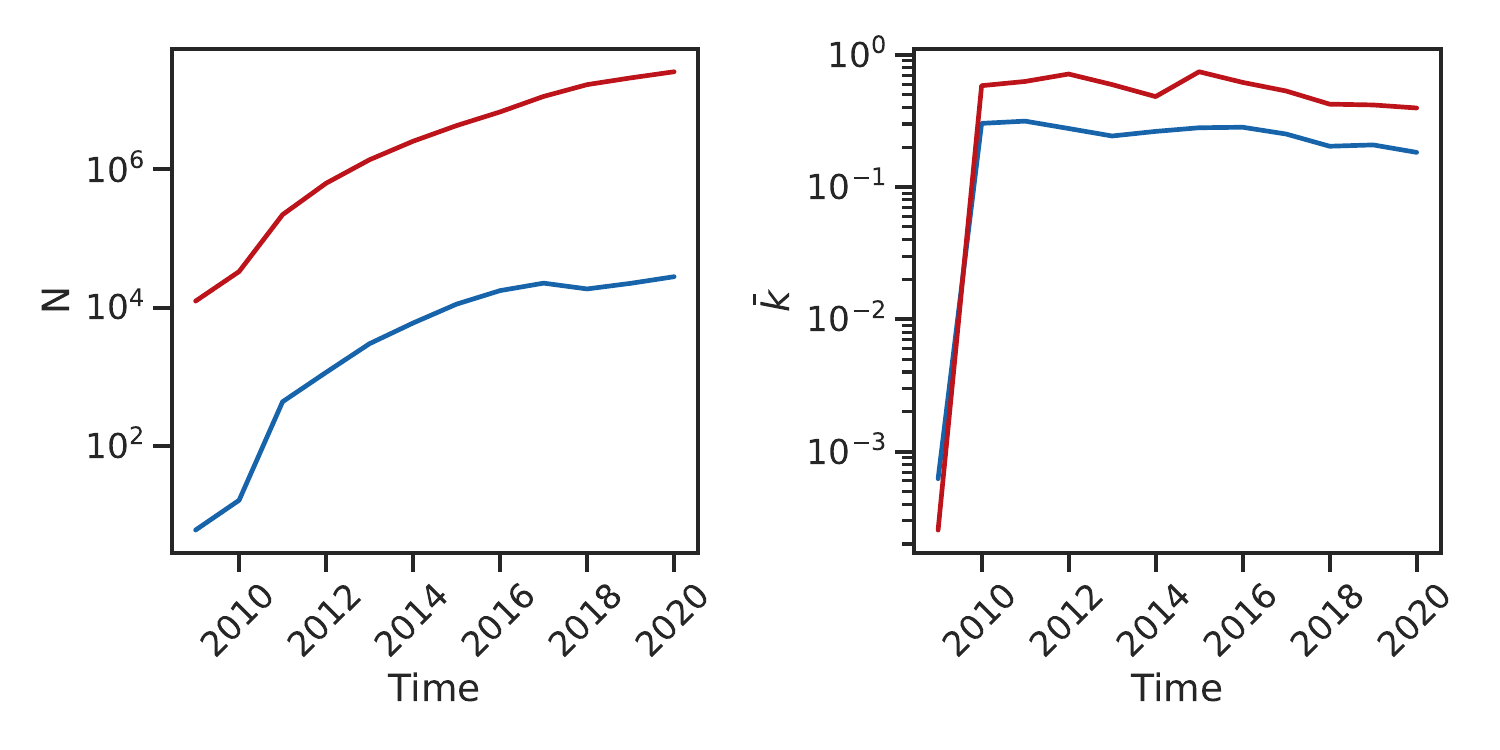}
    \caption{Time evolution of snapshots and stream graph properties. The top panels are related to snapshot representations for all available time resolutions, and the bottom panels are to stream graphs. In the case of snapshots, the left panel presents evolution in terms of the number of nodes (red lines) and the number of edges (blue lines). In the right panel, the average out-degree evolution is shown. In both top panels, the line type and color brightness represent the time resolution, from yearly (solid line), monthly (dashed line), daily (dotted line), and hourly (dots) resolutions. The colors are darker for the higher resolution and brighter for the finest ones. In the case of stream graphs, the left panel shows the evolution of the total node contribution (red) and the average number of transactions at the hour level. The right panel shows the evolution of the average out- and in-degree as defined for stream graphs.}
    \label{fig:snapstream}
\end{figure}

Regarding statistics on the graph's nodes and edges, both temporal representations show the rapid burst of Bitcoin activity during the 2010-2012 period (see Fig.~\ref{fig:snapstream} left panels). In the case of snapshots, considering finer time resolutions such as day and hour allows the observation of much more local peaks. These peaks indicate a specific period of high activity.

Since the definition of node degrees in the context of stream graph and snapshot are different, comparing them is not straightforward. In both cases, we observe a rapid increase in average degrees (see Fig.~\ref{fig:snapstream} right panels). In the context of the stream graph, the time evolution of average degrees is stable after 2010-2011. During this stability period, a gap between incoming and outgoing degrees is present, with $\bar{k_{out}}>\bar{k_{in}}$. This gap indicates denser outgoing transactions (spent BTC) by unit of time and sparser incoming transactions (received BTC). While $\bar{k_{in}}$ remains stable from 2011-2020, $\bar{k_{out}}$ evolution shows a period of decreasing from 2011 to 2014. Snapshots' average degree evolution shows three phases. The first is related to an increasing phase from 2009 to 2014, then a stable phase occurs from 2014 to the end of 2016. The last phase depicts a smooth decrease. When finer resolutions are considered, we observe that the stability period ends with a large peak in July 2015 which can be due to historical Bitcoin events.
\newline

\begin{figure}[h]
    \centering
    \includegraphics[width=\textwidth]{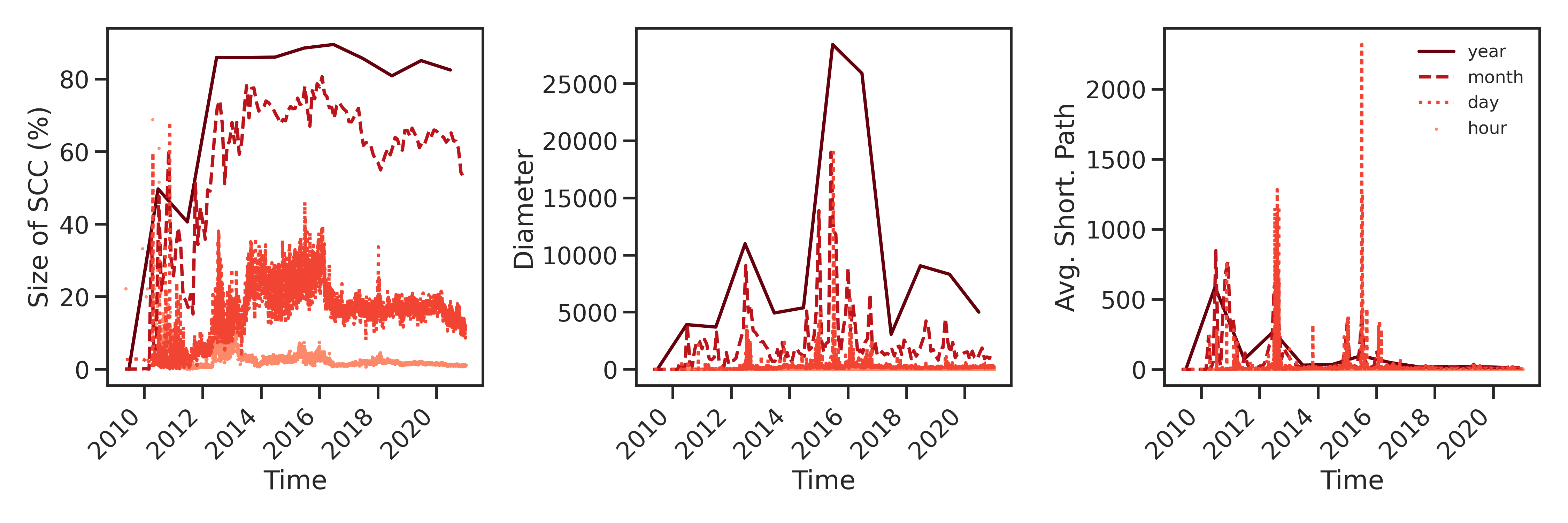}
    \caption{Time evolution of snapshots largest strongly connected component (SCC) properties. The left panel is associated with SCC relative size, the central panel is to diameter, and the right panel is to average shortest path. In every panel, the color brightness depicts the snapshot time resolution, with the darkest colors representing the largest resolutions and the brightest the finest resolution.}
    \label{fig:snapscc}
\end{figure}

As described above, a classical approach to investigate structural changes happening in a sequence of snapshots is to consider the largest SCC, especially the temporal evolution of its relative size, and typical distances. In the case of yearly and monthly snapshots, we observe a rapid increase in the largest SCC relative size up to $80\%$ (resp. $70\%$) in case of year (month) resolution, during the early period of Bitcoin activity, from 2009 to 2013 (see Fig.~\ref{fig:snapscc} left panel). The largest SCC relative size remains stable for the rest of the period captured by the dataset. When we consider finer resolutions, hour and day, the relative size is lower indicating that strong connectivity is a phenomenon emerging at a large time scale.

We observe important structural changes regarding temporal evolution of SCC's diameter and average shortest path (Fig~\ref{fig:snapscc} middle and right panels respectively). In the context of diameter, large peaks appear for both measures. At year resolution, three of them corresponds to the year 2012, the 2015-2016 period, and 2018-2019 period. Regarding the shortes average path, peaks are present for years 2010, 2012 and 2015.
This show that for specific dates, network typical distances are higher indicating longer chain of transactions.
\newline

The results obtained with this analysis show that \textit{ORBITAAL} dataset provides analyzable temporal graphs. The use of standard metrics for snapshots and equivalents for stream graphs show different phase of Bitcoin activity and important structural changes that have happened during the period covered by the dataset. We also show that depending on the time scale of the snapshots, network structures are different, therefore the choice of the time window is important for temporal graph analysis.

\subsection*{Users period of activity}
This last experiment is based on users information data provided by 
\textit{ORBITAAL}. This information include users period of activity and final BTC balance $B$ defined as the difference between the total spent and received BTC.
\subsubsection*{Period of activity}
A user is considered active from its birth date, the first timestamp of activity, to its death date. The death date is either its last transaction timestamp if its final balance is zero, or the last date of the dataset if it keeps a positive balance.

\subsubsection*{Living node}
A living node at time $t$ is any user having a non-zero BTC balance at time $t$. Therefore any node is considered as a living node during its period of activity.
\newline

\begin{figure}[h]
    \centering
    \includegraphics[width=\textwidth]{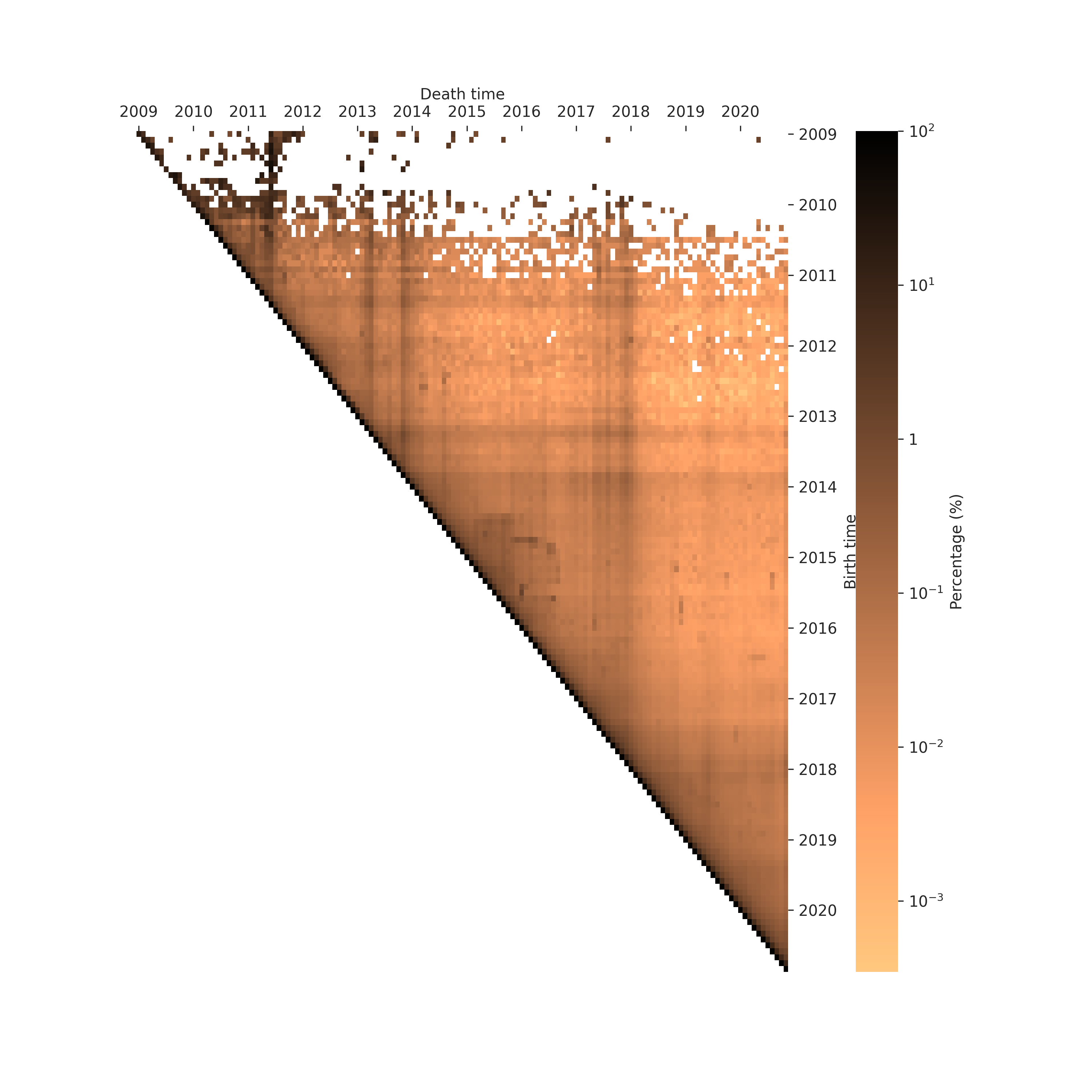}
    \caption{Heatmap representation of users death period. Each row (resp. column) consists in a birth date (death date) at the month resolution. The matrix elements are read such as the $(i,j)$ element gives the proportion of users born at date/row $i$ that died at date/column $j$. Each row sums to one. Color code is logarithmic with darkest cells associated to the period with highest death and brightest cells are associated to low mortality period.}
    \label{fig:usersdeath}
\end{figure}

The user death period heatmap presented in Fig.~\ref{fig:usersdeath} shows that most of the users spent all their BTC a few months after their first activity (short living period). We observe that some dates are marked by higher mortality rates, regardless of their birth date. These dates correspond in particular to the end of 2013, as well as the pre-2018 crisis.

\section*{Usage Notes}

\subsection*{Loading ORBITAAL's dataframe}
\textit{ORBITAAL} dataset contents are data files in \textit{.parquet} and \textit{.csv} formats that can be loaded with Python's packages such as Pandas\cite{pandas} and Pyspark\cite{pyspark}. A readme MarkDown file is also provided with a Quick Start guide containing examples about the dataset usage. 
\subsection*{ORBITAAL's temporal graph analysis}
Due to the large size of stream graphs and snapshots present in \textit{ORBITAAL}, large network analysis packages are required for most analysis. Researchers can use, for instance, SNAP\cite{leskovec2016snap}, Networkit\cite{staudt2016networkit} or Raphtory\cite{raphtory}.
\section*{Code availability}
 
The construction of the dataset can be done easily by following the method given in this article. However, due to the large size of the blockchain, there is a need for high computer performance, especially in terms of CPU, drive memory, and random-access memory (RAM). For this reason, the code used was tailored to our hardware, and thus not shareable in a reproducible way.

\section*{Acknowledgements}

We thank Universite Claude Bernard Lyon 1's support for funding the BIT-STABLENET SENS project.

\section*{Author contributions statement}
R.C. supervised the project, wrote codes for Blockchain data extraction, and address clustering. C.C. wrote codes for temporal graph construction and entity information tables. C.C. performed the data analysis and technical validation. C.C. and R.C. equally contributed to drafting and revising the manuscript.

\section*{Competing interests} 
The authors declare no competing interests.

\section*{Figures \& Tables}

\begin{table}[h]
    \centering
    \begin{tabular}{l|c|c|c}
\cline{1-4}
File name / Archive name & Size in Bytes & Total number of files & Total number of transactions (millions)\\
\cline{1-4}
orbitaal-nodetable.tar.gz&24 GB & 2 & -\\
orbitaal-stream\_graph.tar.gz&24 GB & 13 & 1679\\
orbitaal-snapshot-all.tar.gz&10 GB & 1 & 616\\
orbitaal-snapshot-day.tar.gz&25 GB & 4401 & 1466\\
orbitaal-snapshot-hour.tar.gz&27 GB & 104824 & 1594\\
orbitaal-snapshot-month.tar.gz&23 GB & 145 & 1304\\
orbitaal-snapshot-year.tar.gz&23 GB & 13 & 1258\\
orbitaal-stream\_graph-2016\_07\_08.csv&26 MB & 1 & 0.5\\
orbitaal-stream\_graph-2016\_07\_09.csv&22 MB & 1 & 0.5\\
orbitaal-snapshot-2016\_07\_08.csv&17 MB & 1 & 0.5\\
orbitaal-snapshot-2016\_07\_09.csv&15 MB & 1 & 0.4\\
orbitaal-readme.md & 10.3 KB & 1 & -
    \end{tabular}
    \caption{List of archives in the ORBITAAL repository from Zenodo platform with associated details}
    \label{tab:record1}
\end{table}

\end{document}